\documentclass[a4paper]{jpconf}
\usepackage{graphicx}
\RequirePackage{lscape}
\begin{document}
\title{Small scale direct dark matter search experiments}

\author{Susana Cebri\'an}

\address{Centro de Astropart\'iculas y F\'isica de Altas Energ\'ias (CAPA), Universidad de Zaragoza, 50009 Zaragoza, Spain
\\ Laboratorio Subterr\'aneo de Canfranc, 22880 Canfranc Estaci\'on, Huesca, Spain}

\ead{scebrian@unizar.es}

\begin{abstract}
Experiments based on noble liquids and solid state cryogenic detectors have had a leading role in the direct detection of dark matter. But smaller scale projects can help to explore the new dark matter landscape with advanced, ultra-sensitive detectors based on recently developed technologies. Here, the physics case of different types of small scale dark matter experiments will be presented and many of them will be reviewed, highlighting the detection techniques and summarizing their properties, results and status.
\end{abstract}

\section{Introduction}

In the quest for the dark matter which can be pervading the galactic halo, many different experimental approaches have been followed for its direct detection \cite{review1,review2}. Initiatives for small projects are being now strongly supported by the community \cite{whitepaper}, as they allow to investigate well-motivated candidates with eV-to-GeV masses (Weakly Interacting Massive Particles, WIMPs, and others). The direct detection of dark matter is challenging as it produces a rare signal, concentrated at very low energies with a continuum energy spectrum which appears entangled with background. Therefore, ultra low background conditions and very low energy threshold are a must and the identification of distinctive signatures, like the annual modulation of the interaction rates and the directionality of the signal, would be extremely helpful to assign a dark matter origin to a possible observation. Experiments and projects intended to identify these two distinctive signatures will be presented in Secs.~\ref{secam} and \ref{secdir} while those focused on exploring specifically low mass dark matter and Spin-Dependent (SD) interactions will be discussed in Secs.~\ref{seclm} and \ref{secsd}, respectively; a summary is given in Sec.~\ref{secsum}.

\section{Annual modulation}
\label{secam}

The movement of the Earth around the Sun makes the relative velocity between dark matter particles, in the galactic halo, and a detector, in the Earth, follow a cosine variation producing a time modulation in the dark matter interaction rate $S$, which for each energy bin ($k$) is typically expressed as
\begin{equation}
S_{k}=S_{0,k}+S_{m,k}\cos(\omega(t-t_{0})).
\end{equation}
For a locally isotropic dark matter halo, the modulation has a one year period and a well defined phase, being the maximum rate around June, 2$^{nd}$. The effect should  be weak (at the level of a few percent variation), only noticeable in the low energy region and with a phase reversal at very low energies \cite{drukier,freese1988,freese2013}. No background component is known to mimic the effect, so its observation would be a distinctive signature of the dark matter interaction.

NaI(Tl) crystals read by photomultipliers (PMTs) are being used in projects focused on the detection of the dark matter annual modulation, since, being cheap and robust detectors, it is possible to accumulate a large target and to run for long times in very stable conditions. But new developments have been necessary to get an ultra-low background (reducing for instance $^{40}$K and $^{210}$Pb activity in the crystals) and a low energy threshold. The DAMA/LIBRA experiment operating in the Laboratori Nazionali del Gran Sasso (LNGS) in Italy has collected data over more than twenty years; detectors produced by Saint Gobain company with a mass of 9.7~kg are being used, firstly 9 units and since 2003, 25 detectors. In 2011 all PMTs were replaced, allowing to reduce the software energy threshold from 2 to 1~keV$_{ee}$\footnote{Electron equivalent energy} in the second phase of the experiment. The results from the phase 1 \cite{damaphase1} were confirmed by those of phase 2 \cite{damaphase2}, favoring the presence of a modulation with all the proper features at 12.9$\sigma$ C.L. with an exposure of 2.46~tonne$\cdot$y. The deduced modulation amplitude for the 2-6~keV$_{ee}$ region is $S_{m}=(0.0103\pm0.0008)$~cpd/kg/keV; compatible values were found for different fitting procedures, periods of time, energy regions from 1 to 6~keV$_{ee}$ and detector units. Improved model-dependent corollary analyses after DAMA/LIBRA phase2 have been presented \cite{damacorollary}, applying a maximum likelihood procedure to derive allowed regions in the parameters' space of many different considered scenarios by comparing the measured annual modulation amplitude with the theoretical expectation.

Following the exclusion results presented by many experiments \cite{pdg2018}, there is strong tension when interpreting DAMA/LIBRA annual modulation signal as dark matter, not only in the standard paradigm but even assuming more general halo and interaction models. In addition, no annual modulation effect has been found in experiments using other targets like Xe \cite{xenon100,xmass,lux} or Ge \cite{cdexmod}. In this context, a model-independent proof or disproof with the same NaI target was mandatory, and this is the goal of several projects all over the world. Both COSINE-100 and ANAIS-112 are now in data-taking phase.

COSINE appeared as a joint effort between the KIMS collaboration and the DM-Ice experiment carried out in the South Pole. They operate 8 NaI(Tl) detectors from Alpha Spectra company (106~kg in total) immersed in 2200~l of liquid scintillator at the Yangyang underground Laboratory in South Korea \cite{cosineperformance,cosinebkg}. The Physics run started in September 2016 with a threshold at 2~keV$_{ee}$. From the first 59.5~days of data they have excluded DAMA/LIBRA signal as due to Spin-Independent (SI) WIMPs with a standard halo model \cite{cosinenature}. The first annual modulation analysis corresponding to 1.7~years of data has been presented \cite{cosinemod}; total exposure analyzed is 97.79~kg$\cdot$y as three large crystals were excluded due to low light yield and excessive PMT noise. The best fit modulation amplitude derived for the 2-6~keV$_{ee}$ region is $S_{m}=(0.0083\pm0.0068)$~cpd/kg/keV.

\begin{figure}[]
\includegraphics[width=20pc]{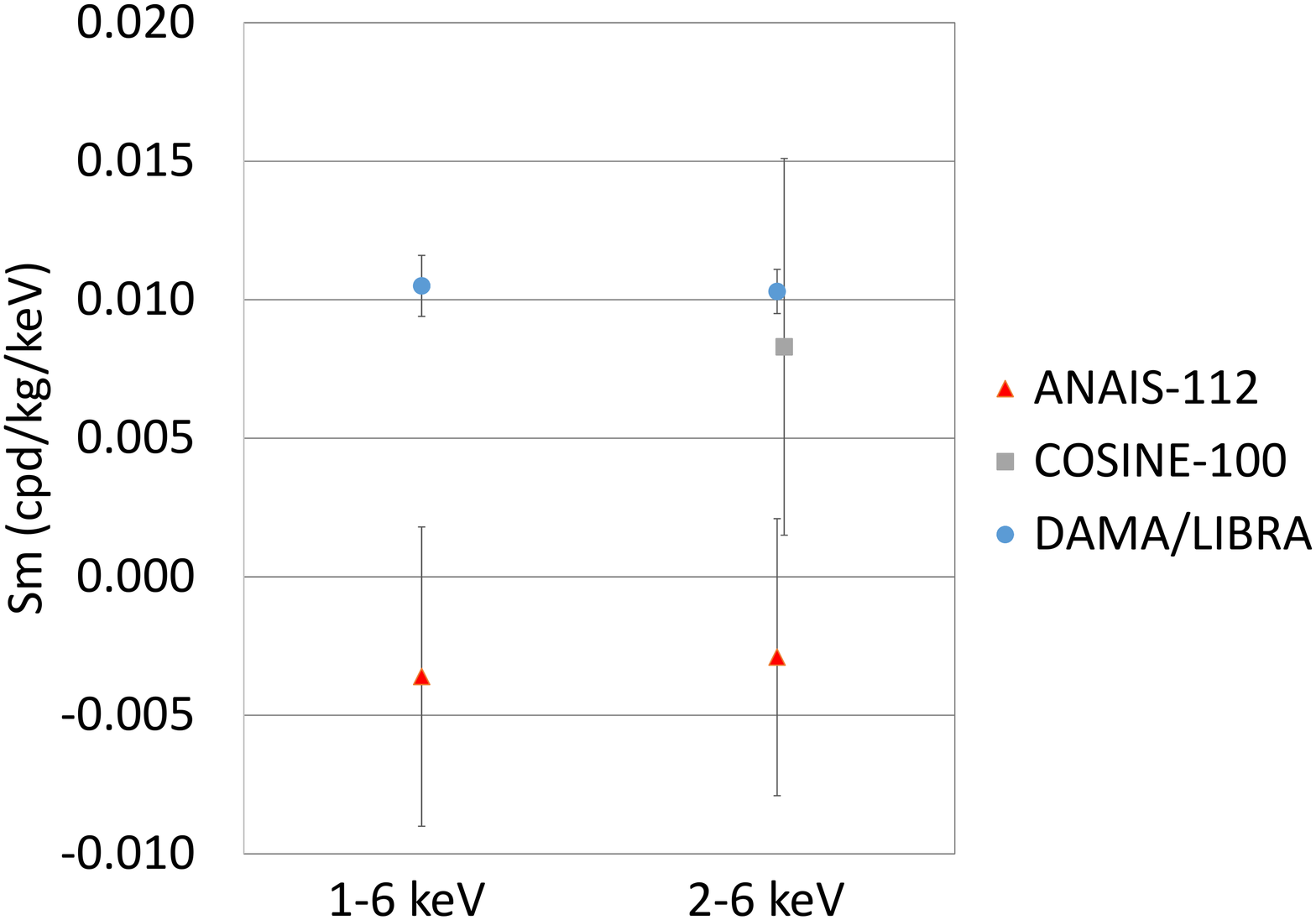}\hspace{2pc}%
\begin{minipage}[b]{14pc}\caption{\label{plotam}Modulation amplitudes obtained by DAMA/LIBRA \cite{damaphase2}, COSINE-100 \cite{cosinemod} and ANAIS-112 \cite{anaistaup} experiments in the energy regions of 1-6~keV$_{ee}$ and 2-6~keV$_{ee}$.}
\end{minipage}
\end{figure}

ANAIS (Annual modulation with NAI Scintillators) is operating 9 NaI(Tl) modules also built by Alpha Spectra (112.5~kg in total) at the Canfranc Underground Laboratory in Spain \cite{anaisperformance,anaisbkg}. The dark matter run is underway since August 2017, with an outstanding light collection of $\sim$15 phe/keV for all modules allowing an energy threshold at 1~keV$_{ee}$. Before any analysis, the sensitivity from the measured background in 1-6~keV$_{ee}$ was evaluated, confirming the possibility to explore the 3$\sigma$ DAMA/LIBRA region in 5 years \cite{anaissensitivity}. The first results on annual modulation for the first 1.5~years of data and exposure of 157.55~kg$\cdot$y have been published \cite{anaismod}; updated results for 2~years were presented at TAUP2019, being the best fit modulation amplitudes  derived for the 2-6~keV$_{ee}$ region $S_{m}=(-0.0029\pm0.0050)$~cpd/kg/keV and for the 1-6~keV$_{ee}$ region $S_{m}=(-0.0036\pm0.0054)$~cpd/kg/keV, incompatible at 2.6$\sigma$ with DAMA/LIBRA results~\cite{anaistaup}. Figure \ref{plotam} compares the modulation amplitudes derived by DAMA/LIBRA, COSINE-100 and ANAIS-112 experiments.

SABRE (Sodium-iodide with Active Background REjection) is in preparation at LNGS \cite{sabre}. It is focused on the development of ultra-high purity NaI(Tl) crystals; a  potassium content of (4.3$\pm$0.2)~ppb has been quantified by ICPMS for a new crystal \cite{sabreK}. They plan to operate $\sim$50~kg and use passive and active (liquid scintillator veto) shielding. Tests with one detector (SABRE Proof of Principle, PoP) are underway with one 3.5~kg crystal shipped from US to Gran Sasso. The main asset of this project is that two identical detectors in northern and southern hemispheres (at LNGS and Stawell Laboratory in Australia) will be implemented to investigate any seasonal effect due to backgrounds, which should show opposite phase.

Also in LNGS, COSINUS (Cryogenic Observatory for SIgnatures seen in Next-generation Underground Searches) pursues a different approach, developing NaI scintillating bolometers~\cite{cosinus}. As the phonon signal is independent of the particle type but the scintillation light is dependent, such a detector has potential to discriminate nuclear recoil events from electronic background. This has been proved with small crystals.

The PICOLON project (Pure Inorganic Crystal Observatory for LOw-energy Neutr(al)ino) is working in Kamioka also in the develoment of highly radiopure NaI(Tl) scintillators after several re-cristallization processes, to reduce $^{40}$K and $^{210}$Pb \cite{picolon}. In the longer term, they plan to install hundreds of kg of NaI(Tl) inside KamLAND.

\section{Directionality}
\label{secdir}

The average direction of the WIMP wind through the solar system comes from the constellation of Cygnus, as the Sun is moving around the Galactic center. A measurement of the track direction of nuclear recoils could be used then to distinguish a dark matter signal from background events (expected to be uniformly distributed) and to prove the galactic origin of a possible signal \cite{spergel,ahlen,mayet}. The reconstruction of tracks is not easy, as they are very short for keV scale nuclear recoils: $\sim$1~mm in gas, $\sim$0.1~mm in solids. It would be desirable to register direction (axis, sense) or at least a head-tail asymmetry by measuring the relative energy loss along the track. A daily modulation of the WIMP direction due to the rotation of Earth would be also observable. There are two approaches for directional detectors: the use of nuclear emulsions and the operation of low pressure ($\sim$0.1~atm) gas targets in TPCs with different electron amplification devices and track readouts, like Multi-Wire Proportional Chambers (MWPC), Micro Pattern Gaseous Detectors (MPGDs) and optical readouts \cite{battat}.

DRIFT (Directional Recoil Identification From Tracks) was the pioneer of directional detectors, using MWPCs attached to a TPC with a large conversion volume (1~m$^{3}$) filled with electronegative gas; in this way, the formed ions (not electrons) are drifted to the readout, to reduce diffusion and optimize track resolution. It operated at Boulby, over more than a decade, using a CS$_{2}$+CF$_{4}$+O$_{2}$ mixture. They measured directional nuclear recoils (from $^{252}$Cf neutrons) quantifying the head-tail asymmetry parameter \cite{battat2} and derived limits for SD WIMP-proton interaction from 54.7~live-days \cite{battat3}. MIMAC (MIcro-tpc MAtrix of Chambers) also operates a dual TPC with a common cathode, but equipped with pixelized bulk Micromegas (micromesh gas structures), at the Modane Underground Laboratory in France since 2012. They work with CHF$_{3}$+CF$_{4}$+C$_{4}$H$_{10}$ and have observed 3D tracks of radon progeny nuclear recoils \cite{riffard}. First observation of $^{19}$F ion tracks at ion beam facilities with angular resolution better than 10$^{o}$ has been reported \cite{tao} and quenching factors of He and F with an ion source in Grenoble have been measured. NEWAGE (NEw generation WIMP search with an Advanced Gaseous tracker Experiment) uses a simplified system with amplification structure and readout in a monolithic detector with a TPC and a micro-pixel chamber. After first operation in surface, long runs at Kamioka in Japan have been made using CF$_{4}$. They have also confirmed the head/tail effect above 100~keV and presented results for SD interaction \cite{nakamura}. A low background detector with polyimide is running since 2018 and new limits have been presented at TAUP2019. DMTPC (Dark Matter Time-Projection Chamber) is based on a TPC equipped with external optical (CCD, PMTs) and charge readouts. Several prototypes have been developed since 2007, operated first at MIT  and then underground at WIPP in US. The measurement of the direction of recoils has been reported too \cite{deaconu}.

Gathering most of the groups working on directional dark matter detection, the CYGNUS collaboration was formed to analyze different gas mixtures, to reduce the energy threshold in these detectors and enlarge the volumes. There are proposals for CYGNUS detectors in different labs in Australia, Italy, Japan, UK and US. CYGNO, working at LNGS, has already operated some prototypes with He/CF$_{4}$ using GEMs, CMOS cameras and PMTs \cite{costa} in the way to build a detector of 1~m$^{3}$; nuclear recoils from a neutron gun with direction and sense visible have been registered in the LEMOn prototype.

An emulsion film made of silver halide crystals dispersed in a polymer can act as target and tracking detector. Nuclear recoils produce nm-sized silver clusters and 3D tracks are reconstructed with an optical microscope. This is the approach followed by NEWSdm (Nuclear Emulsions for WIMP Search with directional measurement) \cite{newsdm}. New generation nuclear emulsions with nanometric grains (NIT (Nano Imaging Tracker) emulsions) have been developed and new fully automated scanning systems overcoming diffraction limits are being prepared; a spatial resolution of 10~nm has been achieved. Test at LNGS with 10~g are underway to assess backgrounds. They propose for a Physics run with 10 kg-year a detector placed on an equatorial telescope (to absorb Earth rotation) to keep orientation towards the Cygnus constellation.

Many other ways to determine the recoil direction have been proposed and are under study, like those based on crystal defect spectroscopy using a diamond target \cite{rajendran}, a DNA	strand detector using biological techniques \cite{dna} or the use of planar targets like graphene (as proposed by PTOLEMY-G \cite{ptolemy}).
			
\section{Low mass dark matter}
\label{seclm}
To probe the proposed dark matter candidates with low masses there are some specific requirements: the use of lighter targets (to keep the  kinematic matching with the dark matter particle), to lower the energy threshold (to detect smaller signals) and even to change the search channel. Light WIMPs cannot transfer sufficient momentum to generate detectable nuclear recoils and therefore scattering off or absorption by both nuclei (NR) or electrons (ER) must be investigated. Different ideas have been proposed to explore different ranges of masses. Here, experiments using semiconductor devices and noble gas detectors will be presented.

CDEX (China Dark matter EXperiment) is using Point-Contact Ge detectors, allowing to reach sub-keV thresholds thanks to a very small capacitance in combination with a large target mass. This was also the approach of the CoGENT detector at Soudan in US \cite{cogent}. Operating at the Jinping underground laboratory in China, in CDEX-1 two detectors of $\sim$1~kg were used reaching an energy threshold of 160~eV$_{ee}$. Limits from an annual modulation analysis have been presented \cite{cdexmod}. In CDEX-10, a 10~kg detector array immersed in liquid N$_{2}$ is being operated and constraints on WIMP-nucleon SI and SD couplings have been presented \cite{cdex10}. Works for CDEX-100 and CDEX-1T are underway too. The TEXONO project uses also an n-type Point-Contact Ge detector \cite{texono}. DAMIC (DArk Matter In CCDs) is based on Si charge-coupled devices (CCDs) where the charge produced in the interaction drifts towards the pixel gates, until readout. In this way, a 3D position reconstruction and an effective particle identification are possible. At SNOLAB in Canada they are operating 7 CCDs with a total mass of 40~g since 2017, achieving a leakage current of 2~e$^{-}$/mm$^{2}$/day and a threshold of 50~eV$_{ee}$. Precise measurements of the quenching factor in Si have been made. Limits of low-mass WIMPs including interaction with electrons and hidden photon dark matter have been presented \cite{damic1,damic2} and new results from an acquired exposure of 13 kg-day will be released soon. For DAMIC-M, to be operated at Modane, more massive CCDs (13.5~g each) will be used, using the Skipper readout: the multiple measurement of the pixel charge allows to reduce noise and achieve single electron counting with high resolution, as already proved. This innovative Skipper readout is also used by SENSEI (Sub-Electron-Noise Skipper CCD Experimental Instrument), working with new generation of CCDs; from first operation at Fermilab in US, constraints on dark matter-electron scattering have been already derived \cite{sensei} and they propose to install a 100~g detector at SNOLAB.

NEWS-G (New Experiments With Spheres-Gas) uses an spherical proportional counter, able to achieve very low energy threshold thanks to very low capacitance ($<$1~pF) for a large volume~\cite{giomataris}. The anode is a small ball at the center of the sphere. The SEDINE detector, consisting of a copper sphere, 60~cm in diameter, filled with Ne-CH$_{4}$ at 3.1~bar (310~g active mass) was operated at Modane. Exclusion results for WIMPs were derived from a 42~day run at 50~eV$_{ee}$ acquisition threshold \cite{newsg1}. A new, larger copper sphere, 140~cm in diameter, has been built; tests have been made in Modane although operation will take place at SNOLAB. The single electron response (gain, drift or diffusion times) has been characterized with a laser system~\cite{newsg2} and quenching factor measurements at Grenoble for H and at TUNL (Duke University) for Ne-CH$_{4}$ are underway. TREX-DM (Tpcs for Rare Event eXperiments-Dark Matter) is based on a gas TPC holding a pressurized gas at 10~bar inside a copper vessel, equipped with the largest Micromegas readouts ever built \cite{trexdmiguaz,trexdmigor,trexdmbkg}. It operates at the Canfranc Underground Laboratory, being presently at the commissioning phase. Runs with mixtures of Argon and Neon with isobutane (atmospheric Ar+1\%iC$_{4}$H$_{10}$ and Ne+2\%iC$_{4}$H$_{10}$) up to 8~bar have been made. Despite some connectivity problems, the preliminary data assessment is positive confirming the prospects for the energy threshold, from 0.4~keV$_{ee}$ down to 0.1~keV$_{ee}$.

\section{Spin-Dependent interaction}
\label{secsd}

Merging PICASSO and COUPP collaborations since 2012, PICO uses bubble chambers, where target liquids are kept in metastable superheated state; sufficiently dense energy depositions then start the formation of bubbles, read by cameras. This detection technique gives no direct measurement of recoil energy but it is almost immune to electronic backgrounds and different targets can be accomodated, most containing $^{19}$F, which offers the highest sensitivity to SD proton couplings. A series of bubble chambers have been operated at SNOLAB. In PICO-60, with 52~kg of C$_{3}$F$_{8}$ a 2.45~keV$_{nr}$ threshold was achieved and the best SD WIMP-proton limit from direct detection has been derived \cite{pico}. Some changes in the design, like the buffer-free concept \cite{bressler}, have been implemented in PICO-40L, already starting the data taking. PICO-500 is a fully funded tonne-scale chamber now in design phase.

\begin{landscape}
\begin{center}
\begin{table}[h]
\centering
\caption{\label{summarytable}Main properties of recent or running small scale dark matter experiments pursuing different goals. Features concerning their detection technique and performance are given, together with the reference to their latest results in the search of dark matter.}
\begin{tabular}{lccccccc}
\br
Experiment & Technique & Laboratory & Target & Mass & Energy Threshold & Future phases & Reference \\
\mr
DAMA/LIBRA & Scintillator & Gran Sasso (Italy) &  NaI(Tl) & 250 kg & 1-2 keV$_{ee}$ & Phase 3 & \cite{damaphase2}\\
COSINE-100 & Scintillator & Yangyang (Korea) & NaI(Tl) & 106 kg & 2 keV$_{ee}$ & & \cite{cosinemod}\\
ANAIS-112 & Scintillator & Canfranc (Spain) & NaI(Tl) & 112 kg & 1 keV$_{ee}$ & & \cite{anaismod} \\
\mr
DRIFT & TPC+MWPC & Boulky (UK) & CS$_{2}$+CF$_{4}$+O$_{2}$ & 0.14 kg & 30-50 keV$_{ee}$ & CYGNUS & \cite{battat3}\\
NEWAGE & TPC+$\mu$PIC & Kamioka (Japan) & CF$_{4}$ & & 50 keV$_{ee}$ &  CYGNUS & \cite{nakamura}\\
\mr
CDEX-10 & Point contact Ge detector & Jinping (China) & Ge & 10 kg &  160 eV$_{ee}$ & CDEX-10X & \cite{cdex10}\\
DAMIC & CCD & SNOLab (Canada) & Si & 40 g & 50 eV$_{ee}$ & DAMIC-M & \cite{damic1}\\
NEWS-G & Spherical Proportional Counter & Modane (France) &  Ne-CH$_{4}$ (0.7\%) & 310 g & 50 eV$_{ee}$ & At SNOLab & \cite{newsg1}\\
\mr
PICO-60 & Bubble chamber & SNOLab (Canada) & C$_{3}$F$_{8}$ &  52 kg &  2.45 keV$_{nr}$ & PICO-40L, PICO-500 & \cite{pico}\\
\br
\end{tabular}
\end{table}
\end{center}
\end{landscape}

\section{Summary and Outlook} \label{secsum}

Other new ideas to directly detect dark matter particles have been proposed and some of them are at the R\&D phase. Scintillating bubble chambers combine the advantages of a bubble chamber with the event-by-event energy resolution of a liquid scintillator; this technique has been established for a 30~g xenon bubble chamber \cite{baxter}. The HeRALD project (Helium Roton Apparatus for Light Dark Matter) proposes the use of superfluid $^{4}$He as target and low temperature calorimeters (TES) as sensors to measure photons and quasiparticles by quantum evaporation \cite{herald}. In the SnowBall proposal supercooled water is used so that an incoming particle triggers crystallization of purified water using a camera for image acquisition; this has been tested with neutrons at $-$20$^{o}$ \cite{snowball}. Lab-grown diamond crystals as target outfitted with cryogenic phonon and charge readout could be sensitive to dark matter scattering of very low mass candidates \cite{kurinsky}. In the so-called paleo-detectors, persistent traces left by dark matter interaction in ancient minerals could be searched for profiting from a large integration time \cite{paleo}.

Small scale experiments looking for the direct detection of dark matter are a very active field, being focused on different physics cases. Table \ref{summarytable} summarizes the main properties of very recent or running experiments of this type. The identification of distinctive signatures of the dark matter interaction, like the annual modulation in the rates or the directionality of the signal, would give a definite confirmation of dark matter detection. Important results (even if still with low significance) have been recently released by NaI(Tl) experiments (COSINE-100 and ANAIS-112) and are in the way to solve the long standing conundrum of the DAMA/LIBRA annual modulation result. Other NaI(Tl) projects, with interesting features, are ongoing too. The construction of a WIMP detector with directional sensitivity is hard, as imaging short tracks with sufficient resolution and sampling large volumes is needed. Low-pressure TPCs with different readouts (MWPCs, Micromegas, $\mu$PICs, optical CCDs, \dots) and the nuclear emulsion technique are being explored. Prototypes of medium size (with volumes from 0.1 to 1~m$^{3}$) have been already built and significant progress on basic requirements (like radiopurity, homogeneity, stability and scalability) is being made. At the same time, there are projects particularly suited to investigate certain interactions and many proposals to explore low mass dark matter by using light targets, achieving extremely low energy thresholds and/or searching for different interaction channels. This has been possible thanks to the development of novel technologies for detectors and sensors, using semiconductor devices, gas detectors and others. All in all, small scale dark matter experiments are already giving very competitive results and more will come in the future.

\ack Special thanks are due to E. Baracchini, A. Chavarria, S. Copello, K. I. Fushimi, G. Gerbier, I.G. Irastorza, I. Katsioulas, C. Krauss, Y. J. Ko, H. Ma, K. Miuchi, T. Noble, P. Privitera, F. Reindl, M. L Sarsa, O. Sato, T. Shutt and M. Yamashita for their help in the preparation of the presentation. I acknowledge funding from Spanish Ministry of Economy and Competitiveness (MINECO) under Grants FPA2016-76978-C3-1-P and FPA2017-83133-P.

\end{document}